
\def\IR{{\hbox{{\rm I}\kern-.2em\hbox{\rm R}}}}
\def\IB{{\hbox{{\rm I}\kern-.2em\hbox{\rm B}}}}
\def\IN{{\hbox{{\rm I}\kern-.2em\hbox{\rm N}}}}
\def\IC{{\ \hbox{{\rm I}\kern-.6em\hbox{\bf C}}}}

\def\IZ{{\hbox{{\rm Z}\kern-.4em\hbox{\rm Z}}}}
\def\to{\rightarrow}
\def\d{{\rm d}}
\def\underarrow#1{\vbox{\ialign{##\crcr$\hfil\displaystyle
{#1}\hfil$\crcr\noalign{\kern1pt
\nointerlineskip}$\longrightarrow$\crcr}}}
%
\def\d{{\rm d}}
\input phyzzx
\tolerance=5000
\overfullrule=0pt

\twelvepoint
\pubnum{IASSNS-HEP-92/53}
\date{August, 1992}
\titlepage
\title{ON BACKGROUND INDEPENDENT OPEN-STRING FIELD THEORY}
\vglue-.25in
\author{Edward Witten
\foot{Research supported in part by NSF Grant
PHY91-06210.}}
\medskip
\address{School of Natural Sciences
\break Institute for Advanced Study
\break Olden Lane
\break Princeton, NJ 08540}
\bigskip
\abstract{A framework for background independent
open-string field theory
is proposed.
The approach involves using the BV formalism -- in a way suggested
by recent developments in closed-string field theory -- to implicitly define
a gauge invariant
Lagrangian in a hypothetical ``space of all open-string world-sheet theories.''
It is built into the formalism that classical solutions of the string
field theory are BRST invariant open-string world-sheet theories and that,
when expanding around a classical solution, the infinitesimal gauge
transformations are generated by the world-sheet BRST operator.    }
\endpage

\chapter{Introduction}

\REF\lovelace{C. Lovelace, ``Stability of String Vacua. 1. A New Picture
of the Renormalization Group,'' Nucl. Phys. {\bf B273} (1986) 413.}
\REF\banks{T. Banks and E. Martinec, ``The Renormalization Group
And String Field Theory,'' Nucl. Phys. {\bf B294} (1987) 733.}
\REF\hughes{J. Hughes, J. Liu, and J. Polchinski, ``Virasoro-Shapiro
{}From Wilson,'' Nucl. Phys. {\bf B316} (1989) 15.}
\REF\periwal{V. Periwal, ``The Renormalization Group, Flows Of
Two Dimensional Field Theory, and Connes' Geometry,'' Comm. Math.
Phys. {\bf 120} (1988) 71.}
\REF\ubanks{T. Banks, ``The Tachyon Potential In String Theory,''
Nucl. Phys. {\bf B361} (1991) 166.}
\REF\bru{R. Brustein and S. de Alwis, ``Renormalization Group Equation
And Non-Perturbative Effects in String Field Theory,'' Nucl. Phys.
{\bf B352} (1991) 451.}
\REF\rol{R. Brustein and K. Roland, ``Space-Time Versus World-Sheet
Renormalization Group Equation In String Theory,'' Nucl. Phys.
{\bf B372} (1992) 201.}
Though gauge invariant open-string and closed-string field theories are now
known, the problem of background dependence of string
field theory has not been successfully addressed.
This problem is fundamental because it is here that one really has
to address the question of what kind of geometrical object the string
represents.  The world-sheet or $\sigma$-model formulation of string
theory is the one known formulation in which anything can be done in
a manifestly background independent way.
It has therefore been widely
suspected that somehow one should do string field theory in the ``space
of all two-dimensional field theories,'' by finding an appropriate gauge
invariant Lagrangian on that space.  The tangent space to the ``space of all
two-dimensional field theories'' should be the space of all local operators,
including operators of very high dimension, time-dependent operators of
negative dimension, and operators containing ghost fields.
This approach, which has been pursued in [\lovelace-\rol],
has two glaring difficulties: (1) because of the ultraviolet difficulties
of quantum field theory, it is hard to define a ``space of all
two-dimensional field theories'' with the desired tangent space (this is
why the sigma model approach to string theory is limited in practice
to a long wavelength expansion);
(2) one has not known what properties such a space should have to enable
the definition of a gauge invariant Lagrangian.

In the present paper, I will propose a solution to the second problem,
for the case of open (bosonic) strings, leaving the first problem to the
future.
Considering open strings means that we consider world-sheet actions of
the form $I=I_0+I'$, where
$I_0$ is a fixed bulk action (corresponding to a choice of closed string
background) and $I'$ is a boundary term representing the open strings.
For instance, the standard closed-string background is
$$I_0=\int_\Sigma \d^2x
\sqrt h\left({1\over 8\pi}h^{ij}\partial_iX^\mu\partial_j
X_\mu +b^{ij}D_ic_j\right). \eqn\abo$$
Here $\Sigma$ is the world-sheet with metric $h$ with coordinates $x^k$,
and $c_i$ and $b_{jk}$ are the usual ghost and antighost fields.
This theory has the usual
conserved BRST current $J^i$.  The corresponding BRST charge
$Q=\oint \d \sigma J^0$ ($\sigma$ is an angular parameter
on a closed-string and ``0'' is the normal direction)  obeys the usual
relations, $$Q^2=0 ~ {\rm  and}~ T_{ij}=\{Q,b_{ij}\},\eqn\aabo$$
with $T_{ij}$ being here the stress tensor.
We then take $I'$ to be an arbitrary boundary interaction,
$$I'=\int_{\partial\Sigma}\d {\sigma} ~~{\cal V}, \eqn\bbo$$
where ${\cal V}$ is an arbitrary local operator constructed from
$X,b,c$; in this paper we consider two ${\cal V}$'s equivalent if they
differ by a total derivative.  A two dimensional theory with action
$I=I_0+I'$, with $I_0$ defined as above and $I'$ allowed to vary,
will be called an open-string world-sheet field
theory.  Our goal will be to define
a gauge invariant Lagrangian on the space of all such open-string
world-sheet theories
(or actually a space introduced later
with some additional degrees of freedom).

\REF\bv{I. A. Batalin and G. A. Vilkovisky,  ``Quantization Of Gauge
Theories With Linearly Dependent Generators,''
Phys. Rev. {\bf D28} (1983) 2567, ``Existence Theorem For Gauge
Algebras,'' J. Math. Phys.
{\bf 26} (1985) 172.}
\REF\stash{J. Fisch, M. Henneaux, J. Stasheff, and C. Teitelboim,
``Existence, Uniqueness, and Cohomology of the Classical BRST Charge
With Ghosts for Ghosts,''
Comm. Math. Phys. {\bf 120} (1989) 379; M. Henneaux and C. Teitelboim,
Comm. Math. Phys. {\bf 115} (1988) 213; J. Stasheff, Bull. Amer. Math.
Soc. {\bf 19} (1988) 287.}
\REF\henn{M. Henneaux,  ``Lectures On The Antifield-BRST Formalism For
Gauge Theories,'' proceedings of the XX GIFT meeting; M. Henneaux
and C. Teitelboim, {\it Quantization Of Gauge Systems}, to be published
by Princeton University Press.}
\REF\ew{E. Witten, ``A Note On The Antibracket Formalism,'' Mod. Phys.
Lett. A {\bf 5} (1990) 487.}
\REF\schw{A. Schwarz, UC Davis preprint (1992).}
\REF\siegel{W. Siegel, ``Covariantly Second-Quantized Strings, II, III,''
Phys. Lett. {\bf 151B} (1985) 391,396.}
\REF\zwiebach{B. Zwiebach, ``Closed String Field Theory: Quantum Action
And The B-V Master Equation,'' IASSNS-HEP-92/41.}
\REF\thorn{C. Thorn, ``Perturbation Theory For Quantized String Fields,''
Nucl. Phys. {\bf B287} (1987) 61.}
\REF\bocc{M. Bochicchio, ``Gauge Fixing For The Field Theory Of The
Bosonic String,'' Phys. Lett. {\bf B193} (1987) 31.}
\REF\wz{E. Witten and B. Zwiebach, ``Algebraic Structures And
Differential Geometry In 2D String Theory,'' IASSNS-HEP-92/4.}
\REF\everlinde{E. Verlinde, ``The Master Equation Of 2D String Theory,''
IASSNS-HEP-92/5.}
This will be easier than it may sound because
the Batalin-Vilkovisky formalism [\bv--\schw] will do much of the work
for us.  The use of this formalism
was suggested by its role in constructing and
understanding classical and quantum closed-string
field theory [\zwiebach], its elegant use in quantizing
open-string field theory [\thorn,\bocc],
and its role in string theory Ward identities [\wz,\everlinde].
In particular, while the BV formalism was first invented for quantizing
gauge invariant classical field theories that are already known,
it was used in closed-string field theory [\zwiebach]
as an aid in finding the
unknown theory; that is how we will use it here.
The BV formalism also has an interesting analogy with the
renormalization group [\bru].

Here is a brief sketch of the relevant aspects of the BV formalism.
(For more information see [\henn].)
One starts with a super-manifold
${\cal M}$ with a $U(1)$ symmetry that we will
call ghost number, generated by a vector field $U$.  The essential
structure on ${\cal M}$
is a non-degenerate fermionic two-form $\omega$ of $U=-1$
which is closed, $\d\omega=0$.  One can
think of $\omega$ as a fermionic symplectic form.  As in
the usual bosonic case, such an $\omega$ has no local invariants; $\omega$
can locally be put in the standard
form $\omega=\sum_a \d \theta_a \d q^a$ with $q^a$ and $\theta_a$ bosonic
and fermionic, respectively.

Just as in the usual case, one can define Poisson brackets
$$\{A,B\}={\partial_rA\over \partial u^K}\omega^{KL}{\partial_lB\over \partial
u^L} \eqn\hbo$$
with $\omega^{KL}$ the inverse matrix to $\omega_{KL}$ and $u^I$ local
coordinates on ${\cal M}$.
(The subscripts $r$ and $l$ refer to derivatives from the right or left.)
These Poisson brackets, which are the BV antibrackets,
obey a graded Jacobi identity.  (At the cost of some imprecision, I will
sometimes refer to $\omega$ rather than the Poisson brackets derived
from it as the antibracket.)
The BV master equation is
$$\{S,S\}=0 \eqn\nbo$$
(which would be vacuous if $\omega$ were bosonic).  An action function
$S$ obeying the master equation is automatically gauge invariant,
with the gauge transformation law
$$\delta u^I=\left(\omega^{IJ}{\partial^2 S\over \partial u^J\partial u^K}
+{1\over 2}{\partial \omega^{IJ}\over\partial u^K}{\partial S\over\partial
u^J}\right)\epsilon^K \eqn\pbo$$
with arbitrary infinitesimal parameters $\epsilon^K$.
It is straightforward to see that $\delta S=\epsilon^K\partial_K\{S,S\}/2
=0$.  (The gauge transformations \pbo\ will only close -- and are only
well-defined, independent of the choice of coordinates $u^I$ -- modulo
``trivial'' gauge transformations that vanish on shell.
These are of the form $\delta u^I=\lambda^{IJ}\partial
S/\partial u^J$, with $\lambda^{IJ}=-\lambda^{JI}$.)

Let ${\cal N}$ be the subspace of ${\cal M}$ on which $U=0$.
We define the
``classical action'' $S_0$ to be the restriction of $S$ to ${\cal N}$.
The classical action has a gauge invariance given, again, by \pbo,
with the $\epsilon^K$ restricted to have $U=-1$.   In
usual applications of the BV formalism to gauge fixing, ${\cal N}$ and
$S_0$ are given,
and the first step is the construction of ${\cal M}$
and $S$ (the latter is required to obey a certain cohomological condition
as well as the master equation).  A general theorem shows that
suitable ${\cal M}$ and ${S}$ exist, but their
actual construction is usually rather painful.  The insight of
Thorn and Boccicchio [\thorn,\bocc] (extending earlier ideas, beginning
with Siegel [\siegel], on the role of the ghosts in string theory) was
that in string theory ${\cal M}$ and ${S}$ are related to ${\cal N}$
and $S_0$ just by relaxing the condition on the ghost number of the fields.
Anticipating this structure was  a help in developing
closed-string field theory, as explained in
[\zwiebach], and will be essential here.

If $S$ is any function, not necessarily obeying the master equation,
one can define a vector field
$V$ by
$$ V^K\omega_{KL} ={\partial_l S\over\partial u^L}.  \eqn\ibo$$
If $S$ has $U=0$, then $V$ has $U=1$.
If we take $S$ as an action functional, then the Euler-Lagrange equations
$0=\d\omega$ are equivalent to $V^I=0$.
As we will see later, the master equation implies that $V^2=0$
or in components
$$V^K{\partial \over\partial u^K}V^I  = 0. \eqn\kubo$$

If we let $i_V$ be the operation of contraction with $V$, then
the definition \ibo\ of $V$ can be written as
$$ i_V\omega =\d S. \eqn\jbo$$
Under an infinitesimal diffeomorphism $u^I\to u^I+\epsilon V^I$ of ${\cal M}$,
a two-form $\omega$ transforms as $\omega\to \omega +\epsilon (i_V\d+\d i_V)
\omega$.
$V$ therefore generates a symmetry of $\omega$ precisely if
$$\left(\d i_V+i_V\d\right) \omega = 0 .       \eqn\jjbo$$
As $\d\omega=0$, this reduces to
$$\d(i_V\omega) = 0, \eqn\nnbo$$
and so is a consequence of
\jbo.  Therefore any vector field derived as in \ibo\ from a function $S$
generates a symmetry of $\omega$.
Conversely, if $V$ is any symmetry of $\omega$, that is any vector
field obeying $\nnbo$, then a function $S$ obeying \ibo\ always
exists at least locally (and is unique up to an overall additive constant).
Possible failure of global existence of $S$ would be analogous
to the multi-valuedness of the Wess-Zumino and Chern-Simons functionals
in field theory.  Since topological questions analogous to this
multi-valuedness would be out of reach at present in string theory, we will
in this paper content ourselves with local construction of $S$.

Suppose that one is given a vector field $V$ that generates a symmetry
of $\omega$ and also obeys $V^2= 0$.  One might wonder if it then
follows that the associated function $S$ obeys the master equation.
This is not quite true, but almost.  The actual situation
is that because of the Jacobi identity of the antibracket, the
map \jbo\ from functions to vector fields is a homomorphism of Lie
algebras; consequently, $V^2$ is the vector field derived from
the function $\{S,S\}/2$, and vanishes precisely if $\{S,S\}$ is constant.

To verify this, one can begin by writing the equation $V^2=0$ in the form
$$ \left[\d i_V+i_V\d,i_V\right] = 0 . \eqn\lbo$$
\jjbo\ then implies that
$$\left(\d i_V+i_V\d\right)i_V\omega = 0. \eqn\mbo$$
Using \nnbo, we get
$$ \d\left(i_Vi_V\omega\right) = 0.    \eqn\inbo$$
This is equivalent to
$$\d\{S,S\}=0, \eqn\obo$$
so that $\{S,S\}$ is a constant, perhaps not zero.
Since this argument can also be read backwards, we have
verified that $V^2=0$ if and only if $\{S,S\}$ is constant.

Looking back at the proof of gauge invariance, we see that the master
equation is stronger than necessary.  A function $S$ obeying \obo\
is automatically gauge invariant, with gauge invariance \pbo.
The generalization of permitting $\{S,S\}$ to be a non-zero constant
is not very interesting in practice, for the following reason.
If we take $S$ to be an action, then the corresponding Euler-Lagrange
equations are $V=0$.  If these equations have at least one solution,
then by evaluating the constant $\{S,S\}$ at the zero of $V$, one finds
that in fact $\{S,S\}=0$.  Therefore, $\{S,S\}$ can be a non-zero
constant only if the classical equations of motion are inconsistent.

I can now explain the strategy for constructing a gauge invariant
open-string Lagrangian.  There are two steps.  (1) On the space of
all open-string world-sheet theories, we will find a fermionic vector field
$V$, of ghost number 1, obeying
$V^2=0$.  (2) Then we will find, on the same space,
a $V$-invariant antibracket, that is, a $V$-invariant fermionic
symplectic form $\omega$ of ghost number $-1$.
The Lagrangian $S$ is then determined (up to an additive constant)
from $\d S=i_V\omega$; it is gauge invariant for reasons explained above.
\foot{On the basis of what happens in field theory, I expect that when
space-time is not compact,
the formula $\d S=i_V\omega$ is valid only for variations of the fields
of compact support; otherwise there are additional
surface terms in the variation of $S$.  Of course, a formula for the
change of $S$ in variations of compact support suffices, together
with locality, to determine $S$ up to an additive constant.}

Of these two steps, the definition of $V$ is straightforward, as we will
see.  The definition of $\omega$ is less straightforward, and a proper
understanding would depend on really understanding what is ``the space of
all open-string  world-sheet
theories.''  I will give only a preliminary, formal
definition of $\omega$.  At least the discussion should serve to make clear
what structures one should want ``the space of all two dimensional
field theories'' to have.

\chapter{Definition Of $V$}

In this paper, our open-string quantum field
theories will be formulated on a disc $\Sigma$.  As one might expect,
this is the relevant case in describing the classical Lagrangian.
The open-string quantum field theories will be required to be invariant
under rigid rotations of the disc, but are not required to have any other
symmetries such as conformal invariance.  That being so,
$\Sigma$ must be endowed with a metric (not just a conformal
structure).  Since rotation invariance will eventually be important,
we consider a rotationally invariant metric on $\Sigma$,
say
$$ \d s^2=\d r^2+f(r)\d\theta^2, ~~~~~0\leq r\leq 1,~~0\leq \theta\leq 2\pi.
            \eqn\mumo$$
The choice of $f$ does not matter; a change in $f$
would just induce a reparametrization of the space
of possible boundary interactions.  In any event,
the metric on $\Sigma$ can be held fixed throughout this paper.

As explained in the introduction,
by an open-string world-sheet
field theory we mean a two dimensional theory with
action $I=I_0+I'$, where
$I_0$ is the fixed bulk action \abo, and $I'$ is a boundary
interaction that does not necessarily conserve the ghost number.
Our first goal in the present section
is to describe an anticommuting vector field, of ghost
number one,  on the
space of such theories.
(Later, in
defining $\omega$, we will add new degrees of freedom to the open-string
field theories.  The construction of $V$ is sufficiently natural that it
will automatically carry over to the new case.)

One way to explain the definition of $V$ is as follows.
An open-string field theory
can be described by giving all possible correlation functions
of local operators in the {\it interior} of the disc.  Thus, the correlation
functions we consider are
$$\langle\prod_{i=1}^n{\cal O}_i(P_i)\rangle  \eqn\dbo$$
with arbitrary local operators ${\cal O}_i$ and $P_i$ in the
interior of $\Sigma$.
The correlation functions \dbo\ obey Ward identities.
Since we choose the $P_i$ to be {\it interior} points, the
Ward identities are entirely
determined by the bulk action $I_0$ of equation \abo\ and are
independent of the choice of boundary contribution in the action.
The boundary interactions determine not the structure of the Ward
identities but the choice of a specific solution of them.
It is reasonable to expect that the space of all solutions of the
Ward identities, for all correlation functions in the interior of $\Sigma$,
can be identified with the space of possible boundary interactions,
since, roughly speaking, the boundary interaction determines how
a left-moving wave incident on the boundary is scattered and
returns as a right-moving wave.  We will use this identification
of the space of solutions of the Ward identities with the space
of open-string theories to define a vector field on the space
of theories.  We also will give an alternative definition that does not use
this identification.

If one is given one solution of the Ward identities, corresponding
to one boundary interaction, then another solution of the Ward
identities can be found by conjugating by any symmetry of the
interior action $I_0$.  An important symmetry is the one generated
by the BRST charge $Q$.  Conjugating by $Q$ is particularly simple
since $Q^2=0$.
If $\epsilon$ is an anticommuting $c$-number, we can form a
one-parameter family of solutions of the Ward identities with
$$\langle\prod_{i=1}^n{\cal O}_i(P_i)\rangle_\epsilon
=\langle\prod_{i=1}^n\left({\cal O}_i(P_i)-i\epsilon\{Q,{\cal O}(P_i)\}
\right)   \rangle.
\eqn\ebo$$

At the tangent space level, this group action on the space of theories
is generated by a vector field $V$, which is anticommuting and
has ghost number 1, since those are the quantum numbers of $Q$,
and obeys $V^2=0$ (or $\{V,V\}=0$) since $Q^2=0$.

Here is an alternative description of $V$.  Let
$J^i$ be the conserved BRST current.  Let $j=\epsilon_{ij}J^i \d x^j$
be the corresponding closed one-form.  Let $C_\alpha$ be a circle
that winds once around all of the $P_i$; for instance, $C_\alpha$ may be
a circle a distance $\alpha$ from the boundary of $\Sigma$, for small
$\alpha$.  Since $j$ is closed,
the contour integral $\oint_{C_\alpha}j$ is invariant under homotopically
trivia
   l
displacements of $C$.
The term in \ebo\ proportional to $\epsilon$ is just
$$\langle \oint_{C_\alpha}j\cdot \prod_i{\cal O}_i(P_i)\rangle,\eqn\fbo$$
as one sees upon shrinking the contour $C_\alpha$ to pick up terms of the
form $\{Q,{\cal O}_i\}$.  On the other hand, we can evaluate \fbo\
by taking the limit as $\alpha\to 0$, so that $C_\alpha$ approaches the
boundary
of the disc.  In this limit, $\oint_{C_\alpha} j$ approaches $\int_{\partial
\Sigma}{\cal V}$ for some local operator ${\cal V}$ defined
on the boundary.  There is no general formula for ${\cal V}$; its
determination depends on the behavior of local operators (in this
case the BRST current) near the boundary of $\Sigma$, and so on
the details of the boundary interaction in the open-string field theory.
But in general, we can interpret $\oint_{\partial\Sigma}{\cal V}$
as a correction to the boundary interaction of the theory, and as such
it defines
a tangent vector field to the space of all open-string field
theories.  This is an alternative description of the
vector field $V$ defined in the previous paragraph.

The correction $\oint_{\partial\Sigma}{\cal V}$ to the boundary
Lagrangian resulted from a BRST transformation of that Lagrangian.
Therefore, ${\cal V}$ vanishes when, and only when, the boundary
interactions are BRST invariant.  The BRST invariant world-sheet
open-string theories are therefore precisely the zeros of $V$.
In other words, the equations
$$V^I=0 \eqn\micoco$$
are the equations of world-sheet BRST invariance.
These equations are certainly background independent in the relevant
sense; no {\it a priori} choice of an open-string background entered
in the construction.
As explained in the introduction, a gauge invariant Lagrangian
with $V^I=0$ as the equations of motion can be constructed provided
we can find a $V$-invariant antibracket on the space of open-string
field theories.

Before undertaking this task, let us make a few remarks about
the relation of the vector field $V$ to BRST invariance.
At a point at which a vector field does not vanish, there is
no invariant way (lacking an affine connection) to differentiate it.
However, at a zero of a vector field, that vector field has
a well-defined derivative which is a linear transformation of
the tangent space.  For instance, if $V$ has a zero at -- say --
$u^K=0$, then we can expand $V^K=\sum_Lq^K{}_Lu^L+O(u^2)$, and
$q^K{}_L$ is naturally defined as a tensor; in fact it can be regarded
as a matrix acting on tangent vectors.  Upon expanding the equation
$V^2=0$ in powers of $u$, one finds that $q^K{}_Lq^L{}_M=0$,
or more succinctly
$$ q^2 = 0. \eqn\fbo$$
In the case of the vector field $V$ on the space of open-string
world-sheet
theories, the tangent space on which the matrix $q$ acts
is the space of local operators that can be added to the boundary
interaction; so it is closely related to the space of first-quantized
open-string states.
Thus essentially $q$ is an operator of ghost number one
and square zero in the open-string Hilbert space; it is in fact
simply the usual BRST operator, for the world-sheet theory
with that particular boundary interaction.

What we have come upon here seems to be the natural off-shell
framework for BRST invariance.  Off-shell one has a vector field
$V$ of $V^2=0$.  $V$ vanishes precisely
on shell, and then the derivative of $V$ is the usual BRST operator
$q$ of $q^2=0$.
In fact, this structure can be seen -- but is perhaps not usually isolated
-- in conventional versions of string field theory.

\chapter{Definition Of The Antibracket}

We now come to the more difficult part of our problem -- defining
the antibracket.  What will be said here is in no way definitive.

It might be helpful first to explain how the antibracket is defined
on shell; see also [\everlinde,\wz].
We start with a conformally invariant and BRST invariant world-sheet theory
with action $I=I_0+I'$, where
$$I'=\int_{\partial\Sigma}\d\sigma \,\,\,{\cal V}, \eqn\poco$$
for some ${\cal V}$.  A tangent vector to the space of classical
solutions of open-string theory is represented by a spin one primary
field  $\delta {\cal V}$.  This perturbation must be BRST invariant
in the sense that
$$\{Q,\delta {\cal V}\}=\d {\cal O} \eqn\mopo$$
for some ${\cal O}$, of ghost number one.
If we are given two such tangent vectors $\delta_i{\cal V},\,\,\,i=1,2$,
then $\{Q,\delta_i{\cal V}\}=\d{\cal O}_i$
for two operators ${\cal O}_i$.  Then we can define the antibracket:
$$\omega(\delta_1{\cal V},\delta_2{\cal V})=\langle {\cal O}_1{\cal O}_2\rangle
 . \eqn\lopo$$
Here $\langle \dots\rangle$ is the expectation value of a product
of operators inserted on the disc, in the world-sheet field theory,
and the ${\cal O}_i$ are inserted at arbitrary points on
the boundary of the disc.
Conformal invariance ensures that the positions at which the ${\cal O}_i$
are inserted do not matter.  With a view, however,
to the later off-shell generalization,
I prefer to write
$$\omega(\delta_1{\cal V},\delta_2{\cal V})=
\oint \d\sigma_1\oint \d\sigma_2\langle {\cal O}_1(\sigma_1){\cal O}_2
(\sigma_2)\rangle
\eqn\loppo$$
with the length element $\d\sigma$ (determined from the
metric on $\Sigma$) now normalized so that the circumference is 1.

The correlation function in \loppo\
is BRST invariant and vanishes if either of the ${\cal O}_i$ is
of the form of $\{Q,\dots\}$, so $\omega$ can be regarded as a two-form on the
space of classical solutions.
$\omega$ has ghost number $-1$ since the ghost number of the vacuum is
$-3$ on the disc, and the shifts $\delta_i{\cal V}\to {\cal O}_i,\,\,i=1,2$
have shifted the ghost number by $+2$.
Non-degeneracy of $\omega$  follows from
its relation
to the Zamolodchikov metric $g(\cdot,\cdot)$
on the space of conformal field theories.
Indeed, if $V$ and $W$ are two spin one primary fields containing no ghost
or antighost fields, and $\delta_1{\cal V}=V$, $\delta_2{\cal V}=\partial
c \cdot W$, then $\omega(\delta_1{\cal V},\delta_2{\cal V})=g(V,W)$.
According to the standard analysis of world-sheet BRST cohomology,
every tangent vector to the space open string solutions can be
put in the form of $\delta_1{\cal V}$ or $\delta_2{\cal V}$.
The non-degeneracy of $\omega$ thus is a consequence of the non-degeneracy
of the Zamolodchikov metric.
$\omega(\cdot,\cdot)$ is really the correct analog of $g(\cdot,\cdot)$
when one includes the ghosts.

In many respects, ${\cal O}$ is
more fundamental than $\delta {\cal V}$.
In string field theory, for instance, the classical string field
is an object of ghost number $1$, corresponding to ${\cal O}$.
At the level of states, the relation between $\delta{\cal V}$ and ${\cal O}$
can be written
$$b_{-1}|{\cal O}\rangle =|\delta {\cal V}\rangle.\eqn\hoho$$
This equation has the following immediate consequence:
$$ b_{-1}|\delta{\cal V}\rangle = 0 . \eqn\ofo$$

I want to reexpress these formulas in terms of operators inserted on the
boundary of the disc (rather than states),
so that they can be taken off-shell.  A
useful way to do this is as follows.  Let $v^i$ be the Killing
vector field that generates a rotation of the disc, and let $\epsilon^j{}_k$
be the complex structure of the disc.
Since $ v$ is a Killing vector field, the operator-valued one-form
$b(v)=v^ib_{ij}\epsilon^j{}_k\d x^k$
is closed.   Let
$$ b_\alpha =\oint_{C_\alpha} b(v)\eqn\ucu$$
where the contour $C_\alpha$ is a distance $\alpha$ from the boundary
of the disc.
Since $b(v)$ is closed, the operator $b_\alpha$, inserted in correlation
functions, is independent of $\alpha$ except when the contour $C_\alpha$
crosses operator insertions.
The  operator $b_\alpha$ acts like $b_{-1}$ on an open
string insertion on the boundary of the disc (it acts as $b_0-\overline b_0$
on a closed string insertion at the center of the disc).
A version of \ofo\ that involves no assumption of conformal or BRST
invariance, and hence makes sense off-shell, is the statement
$$ \lim_{\alpha\to 0}b_\alpha = 0    . \eqn\nurmo$$
This captures the idea that the operators
on the boundary of the disc, which is at $\alpha=0$,
 are annihilated by $b_{-1}$.  A similar version of
\hoho\ that makes sense off-shell is
$$ \lim_{\alpha\to 0}b_\alpha {\cal O}(\sigma)=\delta{\cal V}(\sigma),
 \eqn\urmo$$
with $\sigma$ an arbitrary point on the boundary of the disc.
We will use the symbol $b_{-1}$ as an abbreviation for $\lim_{\alpha\to 0}
b_\alpha$, and so write \urmo\ as $b_{-1}{\cal O}=\delta{\cal V}$.

On shell, when $\delta{\cal V}$ is given, ${\cal O}$
is uniquely determined, either by \mopo\ or by the pair of equations
$$\delta{\cal V}=b_{-1}{\cal O}   \eqn\mmopo$$
and
$$ 0 =\{Q,\cal O\}.
\eqn\mmmopo$$
Off-shell, neither \mopo\ nor \mmmopo\ makes sense.  \mmopo\ still makes
sense, but it does not determine ${\cal O}$ uniquely.  It determines
${\cal O}$ only modulo addition of an operator of the form $b_{-1}(\dots)$.
Actually, since we consider $\delta \cal V$ to be trivial if it is of
the form $\d(\dots)$, ${\cal O}$ is also indeterminate up to addition
of an operator of the form $\d(\dots)$.  The possibility of adding
a total derivative to ${\delta V}$ or ${\cal O}$
causes no problem.  The indeterminacy that causes a problem
is the possibility of adding $b_{-1}(\dots)$ to ${\cal O}$.

We might want to define the antibracket off-shell by the
same formula we used on-shell: $\omega(\delta_1{\cal V},\delta_2{\cal V})
=\oint \d\sigma_1\oint \d\sigma_2\langle {\cal O}_1(\sigma_1)
{\cal O}_2(\sigma_2)\rangle$.  But this
formula is ambiguous, since the ${\cal O}$'s are not uniquely
determined by the $\delta {\cal V}$'s.
I will make a proposal, though far
from definitive, for solving this problem.

\section{The Enlarged Space Of Theories}

By comparison to string field theory, it is easy to see the origin
of the problem.  In string field theory, the basic field is an object
of ghost number 1 -- an ${\cal O}$, in our present terminology --
and the antibracket is defined, accordingly, by a two point function of
${\cal O}$'s.   Since the perturbation of the (boundary term in the)
Lagrangian of the two dimensional field
theory is defined by $\delta{\cal V}=b_{-1}{\cal O}$, in passing from ${\cal
O}$
to $\delta{\cal V}$, we are throwing away some of the degrees of freedom,
namely the operators annihilated by $b_{-1}$.  To solve the problem, one
must find a role in the formalism for those operators.  I will simply
include them by hand.

Instead of saying that
the basic object is a world-sheet Lagrangian of the form
$$I=I_0+\int_{\partial\Sigma}\d\sigma\,\,\,{\cal V}, \eqn\koko$$
I will henceforth say that the basic object is such a world-sheet
Lagrangian together with a local operator ${\cal O}$ such that
$$ {\cal V}=b_{-1}{\cal O}. \eqn\uru$$
The left hand side is now ${\cal V}$, not $\delta{\cal V}$, so we are
changing the meaning of ${\cal O}$.
Since ${\cal V}$ is determined by ${\cal O}$, we can consider the basic
variable to be ${\cal O}$, just as in string field theory.
(However, just as in string field theory, one defines
the statistics of the field
to be the natural statistics of of ${\cal V}$, and the opposite
of the natural statistics of ${\cal O}$.)
Now we can define the antibracket:
$$\omega(\delta_1{\cal O},\delta_2{\cal O})=\oint\d\sigma_1\oint \d
\sigma_2\langle\delta_1{\cal O}(\sigma_1)\delta_2{\cal O}(\sigma_2)\rangle.
\eqn\hogo$$

To formally prove that $\d\omega=0$, one proceeds as follows.
First of all, if $U_i(\sigma_i)$ are any local operators inserted
at points $\sigma_i\in\partial\Sigma$, then
$$ 0 =\langle b_{-1}\bigl(U_1(\sigma_1)\dots U_n(\sigma_n)\bigr)\rangle
\eqn\hodoc$$
This is a consequence of the fact that (as all the operator insertions
are on $\partial\Sigma$),
the correlation function $\langle b_\alpha\cdot\prod_iU_i(\sigma_i)\rangle$
is independent of $\alpha$.  Taking the limit as the contour $C_\alpha$
shrinks to a point, this correlation function vanishes; taking it to approach
$\partial\Sigma$,
we get \hodoc. This Ward identity can be written out in more detail as
$$\eqalign{
0 &=\langle \bigl(
b_{-1}U_1(\sigma_1)\bigr) U_2(\sigma_2)\dots U_n(\sigma_n)\rangle
-(-1)^{\eta_1}\langle U_1(\sigma_1)\bigl(b_{-1}U_2(\sigma_2)\bigr)
\dots U_n(\sigma_n))
\cr &+(-1)^{\eta_1+\eta_2}
\langle U_1(\sigma_1)U_2(\sigma_2)\left(b_{-1}U_3(\sigma_3)\right)
\dots\rangle
\rangle \pm \dots = 0,\cr} \eqn\huccu$$
with $\eta_i$ such that $(-1)^{\eta_i}$ is $\mp 1$ for $U_i$ bosonic
or fermionic (and $\pm 1$ for $b_{-1}U_i$ bosonic or fermionic).
Now if ${\cal O}={\cal O}_0+\sum_it_i{\cal O}_i$, then
$$\d\omega(\delta_i{\cal O},\delta_j{\cal O},\delta_k{\cal O}\rangle
={\partial\over\partial t_i}\omega(\delta_j{\cal O},\delta_k{\cal O})
\pm {\rm cyclic~permutations}.\eqn\alsoo$$
Also, since $\partial/\partial t_i$ is generated by an insertion
of $\delta_i{\cal V}=b_{-1}\delta_i{\cal O}$, we have
$${\partial\over\partial t_i}\omega(\delta_j{\cal O},\delta_k{\cal O})
=\oint\d\sigma_1\,\,\d\sigma_2\,\,\d\sigma_3  \langle
\bigl(b_{-1}{\delta_i {\cal O}}(\sigma_1)\bigr)
\cdot \delta_j{\cal O}(\sigma_2)
\cdot \delta_k{\cal O}(\sigma_3)\rangle. \eqn\balsoo$$
Combining these formulas, we see that $\d\omega=0$ is a consequence
of \huccu.
To establish BRST invariance of $\omega$, one must show that
$\d(i_V\omega)=0$, or in other words that
$$0={\partial\over\partial t_i}\oint\d\sigma_1\d\sigma_2\langle
\delta_j{\cal O}(\sigma_1)\cdot\{Q,{\cal O}\}(\sigma_2)\rangle
\pm i\leftrightarrow j. \eqn\omigo$$
This is proved similarly, using the additional facts that  $\{b_{-1},Q\}=v^i
\partial_i$ (the operator that generates the rotation of the circle) and
and $\oint\d\sigma\,\, v^i\partial_i{\cal O}=0$.

\section{Critique}

What is unsatisfactory about all this?  To begin with, we have been working
formally in a ``space of all open-string world-sheet theories,'' totally
ignoring the ultraviolet divergences that arise when one starts
adding arbitrary local operators (perhaps of very large positive
or negative dimension) to the boundary action.
Even worse, in my view, we have tacitly accepted the view that
a theory is canonically determined by its Lagrangian, in this case
$I=I_0+\int_{\partial\Sigma}\d\sigma \,\,\,{\cal V}$.
That is fine for cutoff theories with a particular cutoff in place,
but runs into difficulties when one tries to remove the cutoff.
In the limit in which one removes the cutoff, the theory really depends
on both ${\cal V}$ and the cutoff procedure that is used.

In our construction, can we work with a cutoff theory or do we need
to remove the cutoff?  The ingredients we needed were rotation invariance,
invariance under $b_{-1}$, and $Q$ invariance.  There is no
problem in picking a cutoff (such as a Pauli-Villars regulator in the
interior of the disc) that preserves the first two (with a modified
definition of $b_{-1}$), but there is presumably no cutoff that preserves
$Q$.  Therefore, we need to take the limit of removing the cutoff.
With a cutoff
in place, one can use the above procedure to define $\omega$ and prove
$\d\omega=0$, but the cutoff $\omega$ will not be BRST invariant; one
will have to hope to recover BRST invariance of $\omega$ in the
limit in which the  cutoff is removed.

This may well work, if a ``space of all world-sheet theories''
(with the desired tangent space) does exist.
The main point that arouses skepticism is actually
the existence of the wished-for
theory space.  Even if such a space exists, there is something
missing (even at a formal level) in my above definition of $\omega$.
Because of the cutoff
dependence at intermediate stages, an open-string field theory does
not really have a naturally defined local operator ${\cal V}$ representing
the boundary interaction.  Even formally, there is some work
to be done to explain what type of objects ${\cal V}$ and ${\cal O}$ are
(independently of the particular cutoff procedure) such that the key
equation ${\cal V}=b_{-1}{\cal  O}$ makes sense.  If this were
accomplished, one could perhaps give a direct formal definition of $\omega$
manifestly independent of cutoff procedure.

\chapter{Conclusions}

I hope that I have at least demonstrated in this paper that in trying
to make sense of the ``space of all open-string world-sheet
field theories,''
the important structure that this space
should possess is
a BRST invariant antibracket.  This will automatically lead to a natural,
background independent open-string field theory in which classical
solutions are BRST invariant world-sheet theories, and on-shell gauge
transformations are generated by the world-sheet BRST operator.
The reasons for hoping that the appropriate antibracket exists are that
it exists on-shell, it exists in string field theory, and it would exist
(as we saw in the last section) if one could totally ignore ultraviolet
questions.  Moreover, the antibracket is the one important structure that
always exists in (appropriate) gauge fixing of  classical field theory.
Other structures, such as metrics in field space,
etc., may or may not exist, but have no general significance in off shell
classical field theory.

Perhaps it is worth mentioning that although our considerations may
appear abstract, they can be made concrete to the extent that one can
make sense of the space of open-string field theories.  One does
not even need the space of {\it all} open-string field theories, since
the considerations of this paper are local in theory space and never involve
sums over unknown degrees of freedom.  If one understands
any concrete family of two-dimensional field theories, one can determine
the function $S$ on the parameter space of this family (up to an additive
constant) by integrating the formula $V^I\omega_{IJ}=\partial_JS$; this formula
can be made entirely concrete (in terms of correlation functions in the given
class of theories).
I hope to give some examples of this elsewhere.

It seems reasonable to expect that a natural antibracket also
exists on the space of all two-dimensional closed-string field theories.
It would be nice to understand at least a formal definition
(even at the imprecise level of \S3).  As for defining an anticommuting
vector field on the space of closed-string theories, I hope that this can
be done by embedding the two-dimensional world-sheet as a non-topological
defect in a topological theory of higher dimension, and using the higher
dimensional world much as we used the disc in the present paper.  Background
independent closed string field theory may therefore be closer than it
appears.

\ack{I would like to thank G. Segal and B. Zwiebach for discussions.}

\refout
\end